\documentclass[twocolumn]{aa}

\usepackage{graphicx}
\usepackage{epstopdf}
\usepackage[intlimits,sumlimits]{amsmath}
\usepackage{deluxetable}
\usepackage{times,epsfig} 
\usepackage{amssymb}
\usepackage{mathrsfs}  

\usepackage{natbib}
\usepackage{txfonts}

\bibpunct{(}{)}{;}{a}{}{,} 
\usepackage{caption}

\usepackage{amsmath}
\usepackage[english]{babel}
\usepackage{longtable}
\usepackage{afterpage}
\usepackage{url}
\usepackage{hyperref}
\usepackage{xcolor}
\definecolor{dark-red}{rgb}{0.4,0.15,0.15}
\definecolor{dark-blue}{rgb}{0.15,0.15,0.4}
\definecolor{medium-blue}{rgb}{0,0,0.5}
\hypersetup{
    colorlinks, linkcolor={dark-red},
    citecolor={dark-blue}, urlcolor={medium-blue}
}

\newcommand{\beqa}{\begin{eqnarray}} 
\newcommand{\eeqa}{\end{eqnarray}}

\newcommand{\bsub}{\begin{subequations}}
\newcommand{\esub}{\end{subequations}}
\newcommand{\beal}{\begin{align}}
\newcommand{\ealn}{\end{align}}

\newcommand{\msun}{M$_{\odot}$}

\newcommand{\brad}{$R_{\mathrm{BB}}$}
\newcommand{\btemp}{$T_{\mathrm{BB}}$}

\newcommand{\Ni}{\ensuremath{^{56}\mathrm{Ni}}}

\begin{document}

\title{The rise and fall of the Type Ib supernova iPTF13bvn}
\subtitle{Not a massive Wolf-Rayet star}

\author{C.~Fremling\inst{1} \and 
J.~Sollerman\inst{1} \and 
F.~Taddia\inst{1} \and 
M.~Ergon\inst{1} 
\and S. Valenti\inst{2,3} 
\and
I.~Arcavi\inst{2,4} \and
S.~Ben-Ami\inst{5} \and
Y. Cao\inst{6} \and
S.~B.~Cenko\inst{7,8}  \and
A.~V.~Filippenko\inst{9} \and
A.~Gal-Yam\inst{5} \and
D.~A.~Howell\inst{2,3}
}

\institute{Department of Astronomy, The Oskar Klein Center, Stockholm University, AlbaNova, 10691 Stockholm, Sweden\and
Las Cumbres Observatory Global Telescope Network, 6740 Cortona Dr., Suite 102, Goleta, CA 93117, USA\and
Department of Physics, University of California, Broida Hall, Mail Code 9530, Santa Barbara, CA 93106-9530, USA\and
Kavli Institute for Theoretical Physics, University of California, Santa Barbara, CA 93106, USA\and
 Benoziyo Center for Astrophysics, The Weizmann Institute of Science, Rehovot 76100, Israel \and
Cahill Center for Astrophysics, California Institute of Technology, Pasadena, CA 91125, USA 
\and
Astrophysics Science Division, NASA Goddard Space Flight Center, Mail Code 661, Greenbelt, MD 20771, USA \and
Joint Space Science Institute, University of Maryland, College Park, MD 20742, USA\and
 Department of Astronomy, University of California, Berkeley, CA 94720-3411, 
USA
}
\date{Received; Accepted}

\abstract
{We investigate iPTF13bvn, a core-collapse (CC) supernova (SN) in the nearby spiral galaxy NGC 5806. This object was discovered by the intermediate Palomar Transient Factory (iPTF) very close to the estimated explosion date and was classified as a stripped-envelope CC~SN, likely of Type Ib. Furthermore, 
a possible progenitor detection in pre-explosion \textit{Hubble Space Telescope (HST)} images was reported, making 
this the only SN~Ib with such an identification. Based on the luminosity and color of the progenitor candidate, as well as on early-time spectra and photometry of the SN, it was argued that the progenitor candidate is consistent with a single, massive Wolf-Rayet (WR) star.}
{We aim to confirm the progenitor detection, to robustly classify the SN using additional spectroscopy, and to investigate if our follow-up photometric and spectroscopic data on iPTF13bvn are consistent with a single-star WR progenitor scenario.} 
{We present a large set of observational data, consisting of multi-band light curves (\textit{UBVRI}, $g^{\prime} r^{\prime} i^{\prime} z^{\prime}$) and 
optical spectra.
We perform standard spectral line analysis 
to track the evolution of the SN ejecta. We also construct a bolometric light curve and perform hydrodynamical calculations to model this light curve to constrain the synthesized radioactive nickel mass
and the total ejecta mass of the SN. Late-time photometry is analyzed to constrain the amount of oxygen.
Furthermore, image registration of pre- and post-explosion \textit{HST} images is performed.}
{Our \textit{HST} astrometry confirms the location of the progenitor candidate of iPTF13bvn, and follow-up spectra securely classify this as a SN~Ib. We use our hydrodynamical model
to fit the observed bolometric light curve, estimating 
the total ejecta mass to be 1.9~\msun\ and the radioactive nickel mass to be 0.05~\msun. The model fit requires the nickel synthesized in the explosion to be highly mixed out in the ejecta. We also find that the late-time nebular $r^{\prime}$-band luminosity is not consistent with predictions based on the expected oxygen nucleosynthesis in very massive stars.}
{We find that our bolometric light curve of iPTF13bvn is not consistent with the previously proposed single massive WR-star progenitor scenario. The total ejecta mass and, in particular, the late-time oxygen emission are both significantly lower than what would be expected from a single WR progenitor with a main-sequence mass of at least 30~\msun.}


\maketitle
\section{Introduction}
\label{sec:intro}

Type Ibc core-collapse supernovae (CC~SNe) have either had their envelopes stripped of hydrogen (SNe~Ib) or stripped of both hydrogen and helium in the case of SNe~Ic \citep[e.g.,][]{1997ARA&amp;A..35..309F}. 
The mass loss could either be due to an extensive wind in a single massive star \citep{2014A&amp;A...564A..30G}, or due to binary interaction \citep{Yoon:2010aa}.

The discovery of iPTF13bvn was made by the intermediate Palomar Transient Factory (iPTF) \citep{Law:2009aa} in the nearby\footnote{$d = 22.5$~Mpc, $\mu = 31.76$~mag \citep{Tully:2009}.} galaxy NGC 5806 on 2013 June 16.24 (UT), just 0.57 days past the estimated explosion date \citep[JD 2456459.17;][]{2013ApJ...775L...7C}. Early-time spectra indicated a likely Type Ib classification. Furthermore, \cite{2013ApJ...775L...7C} used adaptive optics (AO) images and 
{\it Hubble Space Telescope (HST)} pre-explosion archival images to show that there is a possible progenitor within 80 milliarcsec (mas) of the estimated location of the explosion. The luminosity and colors of this progenitor candidate are consistent with those of a single Wolf-Rayet (WR) star. \cite{2013ApJ...775L...7C} further argue that the early-time light curves (LCs) and spectra in the optical and near-infrared, along with the mass-loss rate estimated from radio observations, are all consistent with a WR star as the progenitor.

WR stars are very massive, with zero-age main-sequence masses ($M_{\mathrm{ZAMS}}$) easily surpassing 30~\msun~\citep{1981A&amp;A....99...97M,2007ARA&amp;A..45..177C}. These stars have very strong winds, resulting in high mass-loss rates ($\dot{{M}}$) sometimes even exceeding $\dot{{M}}$~$\approx$~$10^{-5}$~\msun~yr$^{-1}$. It is believed that the high mass-loss rate can cause the entire hydrogen envelope  to be expelled before the star undergoes core collapse \citep{Groh:2013ab,2014A&amp;A...564A..30G}. The result of this would then be a stripped-envelope Type Ib supernova.

Following the discovery and the possible progenitor detection, \cite{Groh:2013aa} used their Geneva stellar evolution models \citep{Groh:2013ab,2014A&amp;A...564A..30G} in combination with the radiative transfer code \emph{CMFGEN} \citep{1998ApJ...496..407H} to model iPTF13bvn. They conclude that the possible progenitor candidate detection and the early-time photometry and spectroscopy are compatible with a model where the progenitor of iPTF13bvn is a single WR star with a main-sequence mass of 32 \msun.

In this paper, we expand on the discussion of iPTF13bvn. 
In Sect.~\ref{sec:Observations and Data Reduction} 
we describe our follow-up observations and give details of the data reduction. 
Section~\ref{sec:prog_13bvn} provides a confirmation of the astrometric identification of the {\it HST} progenitor candidate. 
In Sect.~\ref{sec:Photometry} we present the filtered LCs and describe the construction of the bolometric LC of iPTF13bvn. We use semi-analytic arguments based on the model of \cite{1982ApJ...253..785A} and the methodology developed by \cite{Cano:2013aa} to show that the bolometric properties of the SN are not consistent with the progenitor being very massive (i.e., $M_{\rm ZAMS} > 30$~\msun). 
In Sect.~\ref{sec:Spectroscopy} we report our follow-up spectroscopy, confirm the classification of iPTF13bvn as a SN~Ib and provide velocity measurements of the SN ejecta. The latter are used in Sect.~\ref{sec:hydro} together with the hydrodynamical model \emph{HYDE} (Ergon et al. 2014b, in prep.) 
to further constrain the synthesized nickel and ejecta masses of the explosion and the helium-core mass of the progenitor. In Sect.~\ref{sec:Spectroscopy} we also use late-time photometry ($> 200$ days past the explosion) to constrain the amount of oxygen in the ejecta by comparing our data to the detailed nebular modeling by \cite{jerkstrand2014}. These results are also inconsistent with a very massive progenitor.

\section{Observations and data reduction}
\label{sec:Observations and Data Reduction}

The discovery of iPTF13bvn 
was made with the Palomar Oschin Schmidt 48-inch (P48) telescope. 
\cite{2013ApJ...775L...7C} also reported 
follow-up photometry obtained with the P48, the robotic Palomar 60-inch telescope \citep[P60;][]{2006PASP..118.1396C}, and the Las Cumbres Observatory Global Telescope network \citep[LCOGT;][]{2013PASP..125.1031B} up to 20 days past the discovery. Here we present photometry from the same telescopes up to 90 days past discovery, as well as additional data from the Nordic Optical Telescope (NOT) at La Palma obtained around 240 days past discovery.

To produce LCs from the P48, P60, and NOT data, we have used an image-subtraction procedure with templates consisting of images of NGC 5806 obtained approximately one year before the discovery of iPTF13bvn. 
The point-spread function (PSF) is determined and matched prior to image subtraction, and 
is subsequently used for PSF fitting photometry on the subtracted frames. 
For the images obtained by LCOGT we lack templates, and thus we estimate the galaxy contribution by fitting a low-order surface, and then performing PSF fitting photometry. 
We calibrate our P48 and Sloan filter data against a minimum of 
10 SDSS \citep{2014ApJS..211...17A} stars in the field. The Johnson-Cousins {\it UBVRI} filter data were calibrated against Landolt standard stars \citep{1992AJ....104..340L} observed during photometric nights.

Optical and near-infrared spectra of iPTF13bvn were obtained with many different telescopes and instruments, starting
within 24~hours after the discovery. Spectra until 16~days past the discovery were published by \cite{2013ApJ...775L...7C}. In this paper, we present 6 additional later-time optical spectra taken 18--86~days past the SN discovery. These were obtained with the Palomar 200-inch telescope (P200) using the Double Spectrograph \citep[DBSP;][]{1982PASP...94..586O}, with the NOT using the Andalucia Faint Object Spectrograph (ALFOSC), and with the two Keck 10~m telescopes using the Low Resolution Imaging Spectrograph \citep[LRIS;][]{1995PASP..107..375O} on Keck~1 and the Deep Extragalactic Imaging Multi-Object Spectrograph \citep[DEIMOS;][]{2003SPIE.4841.1657F} on Keck~2. 

All spectra were reduced using standard pipelines and procedures for each telescope and instrument. 
All spectral data and corresponding information is available 
via WISeREP\footnote{\href{http://www.weizmann.ac.il/astrophysics/wiserep/}{http://www.weizmann.ac.il/astrophysics/wiserep/}} \citep{Yaron:2012aa}. 
More data and details will be published by Fremling et al. (2014b, in prep.), where we will also investigate the explosion site of iPTF13bvn in more detail.

\begin{figure}
\centering
\includegraphics[width=\columnwidth]{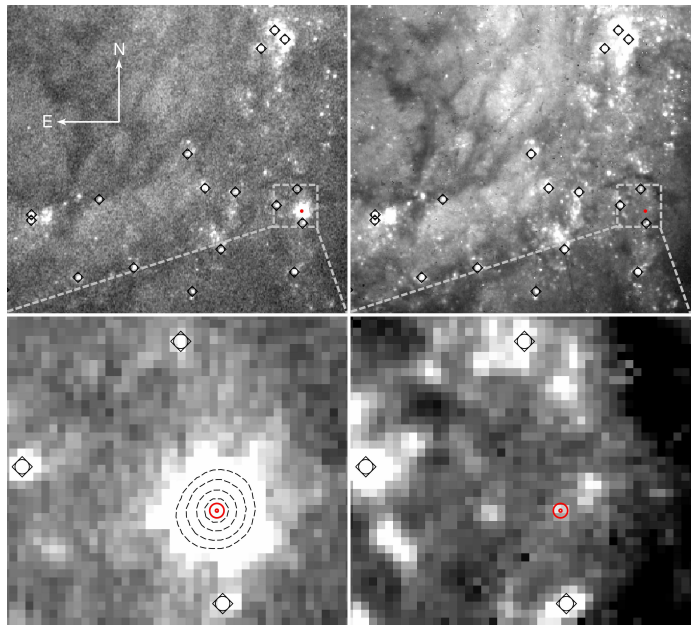}
\caption{Progenitor identification of iPTF13bvn based on registering an {\it HST} WFC3 image of the SN to 
a stacked archival pre-explosion {\it HST} ACS image. The left panels show an image taken when iPTF13bvn was clearly visible, and the panels to the right show the pre-explosion image. The field of view in the upper panels is $16\arcsec\times16\arcsec$.
The location of the SN is marked by the grey dashed square in the upper panels, and the lower panels show this region in detail.
Common point sources used for the registration are marked with boxed black circles. The intensity contours of the bright SN in the lower-left panel are traced with black dashed lines. The centroid of the SN is indicated with red circles in each panel; the smaller and larger red circles respectively represent 1$\sigma$ and 5$\sigma$ uncertainties in the registration. The lower-right panel shows the progenitor identification; the centroid of the SN is directly coincident with one object.}
\label{fig:prog_13bvn}
\end{figure}

\section{Progenitor identification}
\label{sec:prog_13bvn}
A progenitor candidate for iPTF13bvn was
identified in pre-explosion {\it HST} images by \cite{2013ApJ...775L...7C}\footnote{First reported in ATels 5140 and 5152 \citep{2013ATel.5152....1A,2013ATel.5140....1A}.}. This was
done by registering AO images taken with OSIRIS and the
laser-guide-star-AO system on the Keck~1 telescope. \cite{2013ApJ...775L...7C} present a single source
offset by 
about $80\pm40$~mas from the SN position, 
barely within the estimated 2$\sigma$ image-registration uncertainty.
We have used recent archival {\it HST}
(WFC3) images\footnote{Obtained on 2013 Sep. 2.37 (UT), 
GO-12888, PI S. Van Dyk.} of iPTF13bvn to corroborate
this identification. Figure \ref{fig:prog_13bvn} shows the result of
registering a WFC3 image (filter F555W) to a stacked (filters F814W, F555W, and F435W) archival {\it HST} (ACS/WFC) image from
2005 \citep{Smartt:2009} using 25 common point sources.

Our registration procedure is based on standard \emph{MATLAB} functions and is set up as follows. The centroids of the selected common point sources in the images are derived from fits of two-dimensional Gaussians that also allow for rotation; the geometric transformation is subsequently determined as a second-order polynomial transformation with 12 parameters, and the polynomial transformation is finally applied using bicubic interpolation.
Using this procedure we achieve a standard
deviation of 0.18 {\it HST} (ACS/WFC) pixels\footnote{The {\it HST} images used here have a pixel scale of 50 mas.} in the offset between the post-explosion
centroids and the pre-explosion centroids after image registration. 
If we combine this standard deviation with the uncertainty in the
centroid of the SN in the post-explosion image, we find a total
(1$\sigma$) uncertainty of 9~mas in our determination of the location of
iPTF13bvn, corresponding to a projected distance of 1~pc. We find
that the center of the SN explosion is almost exactly coincident 
($+1.7$~mas in $\alpha$, $+3.6$~mas in $\delta$)
with the centroid of the previously proposed progenitor candidate. \cite{2013ApJ...775L...7C} 
reported 
the absolute (Vega) magnitudes for this source:
$M_B=-5.52\pm0.39$, $M_V=-5.55\pm0.39$, and $M_I=-5.77\pm0.41$
mag. These are consistent
with a single WR star \citep{Eldridge:2013aa,Groh:2013aa}.
However, \cite{2013ApJ...775L...7C} also caution that other scenarios (e.g., a binary
system, or a small cluster) can result in similar colors and absolute magnitudes. 

\begin{figure}
\centering
\includegraphics[width=\columnwidth]{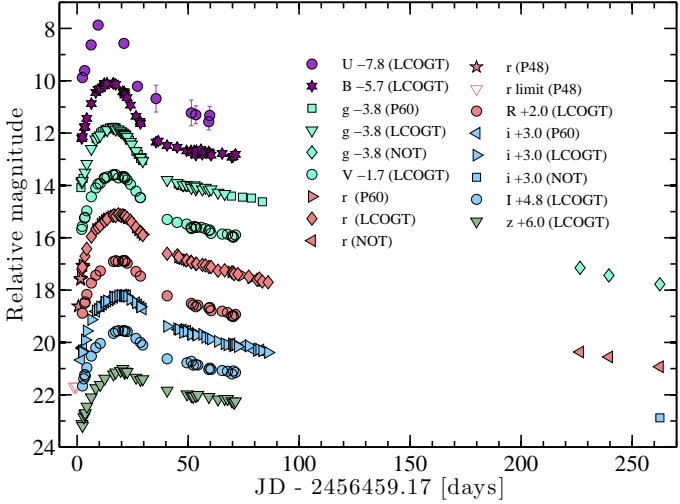}
\caption{Multi-band light curves of iPTF13bvn. Obtained with the P48 and P60 telescopes, the LCOGT telescopes, and the NOT.}
\label{fig:lc_13bvn}
\end{figure}

\section{Light curves}\label{sec:Photometry}

The P48 $r^{\prime}$-band pre-explosion limits
and the optical LCs of iPTF13bvn are displayed in Fig. \ref{fig:lc_13bvn}.  We have constructed a quasi-bolometric LC\footnote{We construct a full bolometric LC and a hydrodynamical model of the LC in Sect.~\ref{sec:hydro}.}
by integrating the flux in the {\it BVRI} bands, and we have determined the black-body (BB) parameters of iPTF13bvn by fitting BBs to the spectral energy distribution
(SED) derived from the full set of {\it UBVRI} and $g^{\prime} r^{\prime} i^{\prime} z^{\prime}$ LCs from the P60 and the LCOGT telescopes. 
For the quasi-bolometric LC calculation, we have interpolated the flux in the {\it BVRI} bands to the
dates of the $V$-band measurements. For these calculations we adopt a distance modulus $\mu = 31.76$~mag \citep{Tully:2009}, a Milky Way (MW) color excess $E(B-V)_{\mathrm{MW}}=0.0278$ mag \citep{2011ApJ...737..103S}, a host-galaxy color excess $E(B-V)_\mathrm{host}=0.0437$ mag \citep{2013ApJ...775L...7C}, and the \cite{1989ApJ...345..245C} extinction law with $R_V=3.1$. 
For the BB fits we have also used weights proportional to the uncertainties in the photometry, and we have interpolated or linearly extrapolated the flux in the {\it UBVRI} and $g^{\prime} r^{\prime} i^{\prime} z^{\prime}$ bands to the dates of both the $V$- and $r^{\prime}$-band measurements. 

\begin{figure}
\centering
\includegraphics[width=\columnwidth]{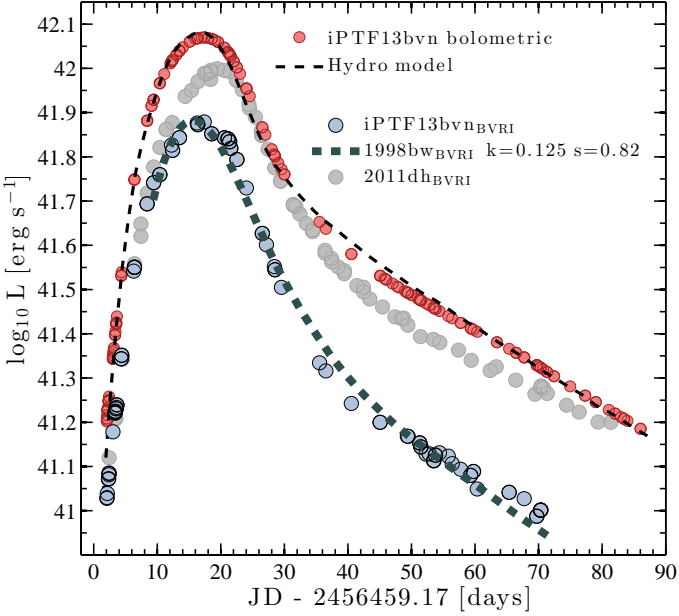}
\caption{Quasi-bolometric {\it BVRI} LC of iPTF13bvn (blue circles), compared to the quasi-bolometric LCs of SN~2011dh (gray circles) and SN~1998bw (thick dashed line). The LC of SN~1998bw has been scaled by $k=0.125$ and stretched by $s=0.82$. The figure also shows the estimated bolometric LC of iPTF13bvn (red circles) compared with our hydrodynamical model of the bolometric LC (thin dashed line).}
\label{fig:bol_lc}
\end{figure}

\begin{figure}
\centering
\includegraphics[width=\columnwidth]{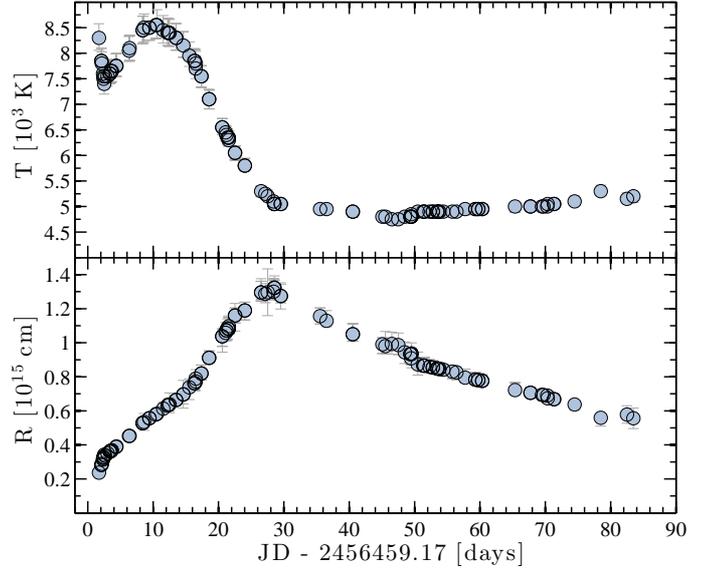}
\caption{Black-body temperature (top panel) and black-body radius (bottom panel) of iPTF13bvn derived from BB fits to the photometry.}
\label{fig:btempbrad}
\end{figure}

The quasi-bolometric LC and the BB parameters
of iPTF13bvn are shown in Fig.~\ref{fig:bol_lc} and Fig.~\ref{fig:btempbrad}, respectively. In Fig.~\ref{fig:bol_lc} we also display
the quasi-bolometric {\it BVRI} LC of the Type IIb SN 2011dh \citep{Ergon:2014aa}, 
as well as a scaled and stretched quasi-bolometric {\it BVRI} LC of
SN 1998bw \citep[SN~Ic-BL;][]{2011AJ....141..163C}. 
We find that the peak {\it BVRI} luminosity of iPTF13bvn 
is somewhat lower than that of SN 2011dh, indicating a slightly lower 
$^{56}$Ni mass\footnote{\Ni\ mass estimates for SN 2011dh are $0.075\pm0.025$~\msun\ by \citet{Ergon:2014aa} and $\sim0.07$~\msun\ by \cite{2013MNRAS.436.3614S}.}.
Furthermore, the width of the LC peak appears to be narrower for iPTF13bvn than for both SNe
1998bw and 2011dh. The narrow peak already disfavors a
massive single-star (WR) progenitor scenario, since such a massive progenitor
(i.e.,~$M_{\mathrm{ZAMS}}~\gtrsim$~30~\msun) should have a large ejecta mass resulting in a
slower evolution of the bolometric LC with a broader peak
\citep{1982ApJ...253..785A}. 

We find that the {\it BVRI} LC of SN 1998bw, multiplied by the scaling
factor $k=0.125$ and stretched by the factor $s=0.82$ using
\begin{equation}
\mathrm{iPTF13bvn}_{L_\mathrm{{BVRI}}}(t)=k \times \mathrm{SN1998bw}_{L_\mathrm{{BVRI}}}(t\times s)
\end{equation}
is very well matched with our {\it BVRI} LC of iPTF13bvn. These parameter values were found by varying $s$ and $k$ and minimizing the residual between the two SN LCs up to 40 days past the estimated explosion dates. 
Interpolating the $s$ and $k$ factors found for the sample of
SNe~Ibc in \cite{Cano:2013aa} provides\footnote{\cite{Cano:2013aa} assume a constant opacity, $\kappa = 0.07$~cm$^2$~g$^{-1}$, in the calculations of the physical parameters of the SNe in their sample.} $M_{\Ni}=0.05\pm0.02$~\msun\ and $M_{\mathrm{ej}}=1.8\pm0.3$~\msun. These interpolations only include SNe with photospheric velocities in the range 7000--11\,000~km~s$^{-1}$, to roughly match the velocities derived from
the early-time spectra of iPTF13bvn in \cite{2013ApJ...775L...7C}, 
and also in later spectra (Sect.~\ref{sec:Spectroscopy}). 
A comparison of our derived masses for iPTF13bvn to
the median nickel mass ($M_{\Ni}=0.15-0.18$~\msun) and ejecta
mass ($M_{\mathrm{ej}}=3.9\pm1.6$~\msun) of the full SN~Ibc sample in
\cite{Cano:2013aa} 
indicates that iPTF13bvn has a lower
$^{56}$Ni mass and also likely a lower
ejecta mass than the average SN~Ibc.

From our black-body fits to the SED of iPTF13bvn we derive the
best-fitting BB radius (\brad) and  temperature
(\btemp) as a function of time. We show the evolution of \brad,
which can be interpreted as a rough approximation of the photospheric
radius, in the bottom panel of Fig.~\ref{fig:btempbrad}. The
best-fitting \btemp\ is shown in the top panel of 
Fig.~\ref{fig:btempbrad}. Both \brad\ and \btemp\  are similar to those of SN 2011dh \citep{Ergon:2014aa},
with \btemp\ peaking at $\sim 8500$~K and \brad\
peaking at $\sim1.4\times10^{15}$~cm for both objects. Compared to the larger
sample of SNe~Ibc presented by \cite{taddia2014}, both
\btemp\ and \brad\ appear to be consistent with those of normal SNe Ibc. If \brad\ 
is linearly extrapolated to \brad~=~0,
we find an explosion date ($t_0$) of 2013 June~$15.75\pm0.3$ (UT),
consistent with the $t_0$ estimate of June 15.67 by \cite{2013ApJ...775L...7C} using a
power-law fit to the early-time $r^{\prime}$-band LC. 

\section{Spectra}
\label{sec:Spectroscopy}

\cite{2013ApJ...775L...7C} suggested a Type Ib classification of iPTF13bvn, partly based on the tentative emergence of \ion{He}{i} features (5876, 6678, 7065 \AA) in optical spectra obtained 10--15~days past explosion. 
Our later-time optical spectra clearly demonstrate the continued emergence of these features (Fig.~\ref{fig:spec_evol}), as well as of the \ion{He}{i} 5016~\AA\ line. At $\sim 35$~days past explosion, we find a very good match between the spectra of iPTF13bvn and those of the Type Ib SNe 2008D \citep{Modjaz:2009aa} and 2007Y \citep{Stritzinger:2009aa} (bottom panel of Fig.~\ref{fig:spec_evol}). Thus, it is clear that the Type Ib spectral classification holds. 

\emph{SYNOW} fits were performed by \cite{2013ApJ...775L...7C} on the early-time spectra of iPTF13bvn, suggesting a photospheric velocity ($v_{\mathrm{ph}}$) of 
10\,000~km~s$^{-1}$
at +3~d and 8000~km~s$^{-1}$
at +12~d. 
To deduce the photospheric velocity, the absorption minimum of the \ion{Fe}{ii} 5169 \AA \  line is likely a more robust tracer \citep{Dessart:aa}.
We detect the absorption minimum from this line in spectra later than +10~d post-explosion, and show the deduced photospheric velocities in Fig.~\ref{fig:line_vel}. 
We find $v_{\mathrm{ph}}\approx$ 10\,000~km~s$^{-1}$ 
at +10~d, in agreement with the 8000--10\,000~km~s$^{-1}$ estimate by \cite{2013ApJ...775L...7C}. Figure~\ref{fig:line_vel} also shows the photospheric velocity deduced from BB fits to the photometry (Sect.~\ref{sec:Photometry}), as well as line velocities from the absorption minimum of 
the \ion{He}{i} 5876, 6678, 7065~\AA\ lines. Typical uncertainties in the velocity estimates are 15\% before 15~d past the explosion and 5--10\% at later times. Up until +10~d the different velocity estimates appear to be in good agreement. At later stages, especially the \ion{He}{i} 5876 \AA\ velocity starts to
deviate, and ceases to be a useful tracer. Similar behavior was observed for SN 2011dh \citep{Ergon:2014aa}. In conclusion, we confirm that a photospheric velocity of 8000--10\,000~km~s$^{-1}$ 
during the LC peak of iPTF13bvn is a reasonable estimate. For our hydrodynamical model of the SN in Sect.~\ref{sec:hydro} we use the \ion{He}{i} 5876~\AA\ line as a tracer up until +10~d and thereafter the \ion{Fe}{ii} 5169 \AA \ line up until +40~d.

\afterpage{
\begin{figure*}
\centering
\includegraphics[width=16cm]{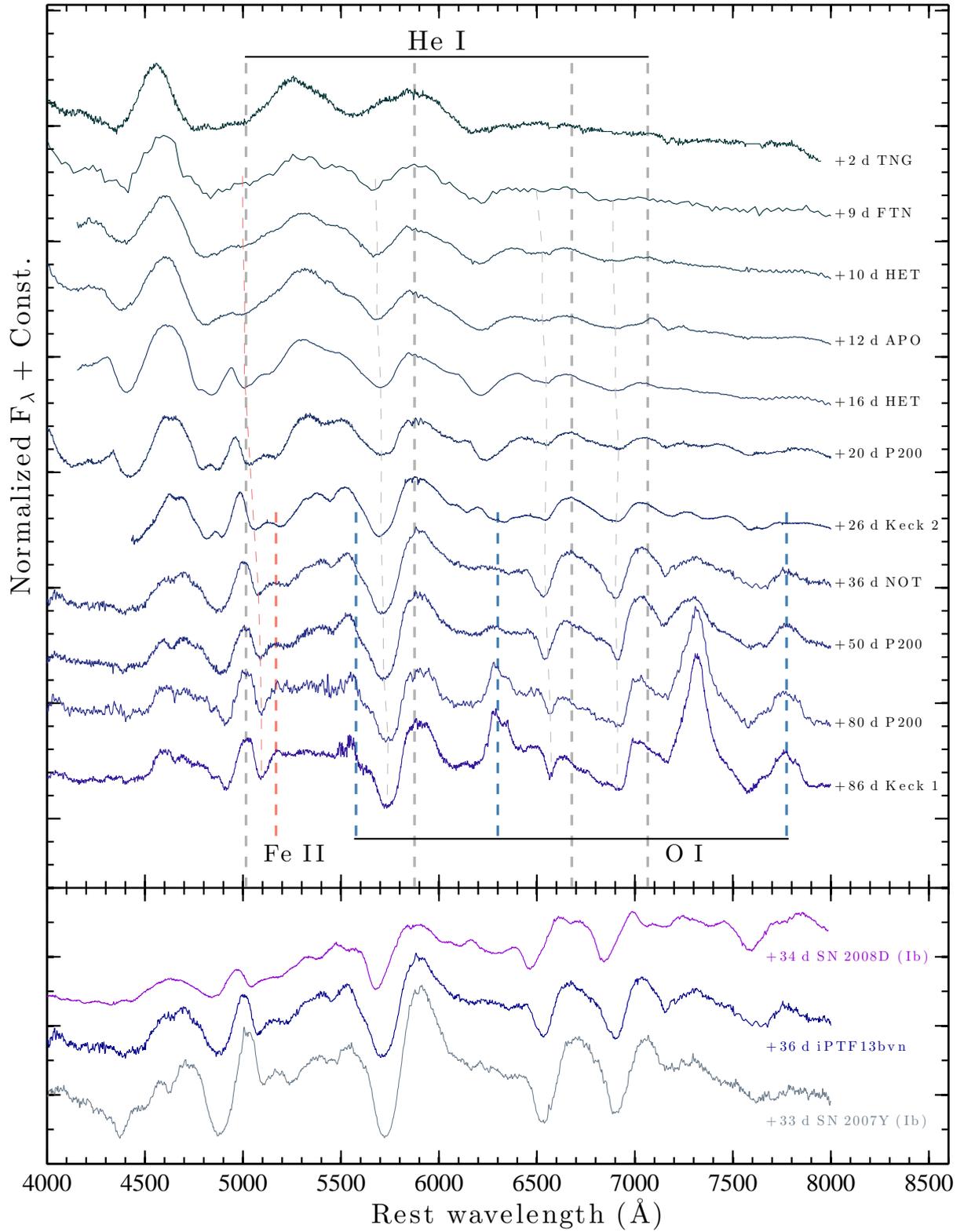}  
\caption{\label{fig:spec_evol}Spectral evolution of iPTF13bvn in the wavelength range 4000--8000~\AA\ (top panel). Comparison of the visible spectra of iPTF13bvn to those of other SNe~Ib at approximately 35~d past the explosion (bottom panel). Thick dashed lines mark the central wavelength of the marked emission lines at rest. Thin dashed lines mark the absorption minima associated with the emission lines derived from Gaussian fits to the absorption features. The spectra up to 16~d past the explosion have been selected from the early spectral sequence presented by \cite{2013ApJ...775L...7C}.}
\end{figure*}
\clearpage
}

\subsection{Constraining the oxygen mass}
\label{sec:oi}
Our optical spectra at epochs later than +50~d show a clear emergence of \ion{O}{i} features (5577, 6300, 6364, and 7774 \AA). For the massive single-star progenitor scenario proposed by \cite{Groh:2013aa}, a large O mass is expected, and our observations allow us to test this. 

The concept here is the realization that more massive stars produce larger amounts of metals, and in particular the oxygen nucleosynthesis is a strong and monotonic function of $M_{\mathrm{ZAMS}}$ \citep{2007PhR...442..269W}. This was exploited by \cite{2014MNRAS.tmp..440J}, who find a strong dependence between the ratio ($O_\mathrm{r}$) of the
[\ion{O}{i}] 6300, 6364~\AA\ line luminosity to the total $^{56}$Co decay power,
and $M_{\mathrm{ZAMS}}$ of SNe IIP \citep[see, e.g., the top panel of Fig. 4 in][]{2014MNRAS.tmp..440J}. 
More recently, \cite{jerkstrand2014} have also modeled 
nebular stripped-envelope CC~SNe,
and a similar strong dependence 
between $O_\mathrm{r}$ and $M_{\mathrm{ZAMS}}$ was found.
Our spectrum obtained at +86~d is likely not yet in the optically thin nebular phase. However, the 
[\ion{O}{i}]~6300,~6364 \AA\  line luminosity we measure at this phase is 
consistent with that of SN~2011dh, which
likely was the result of the explosion of a 12--13~\msun\ $M_{\mathrm{ZAMS}}$ 
star with an oxygen mass of 0.3--0.5~\msun\ at the time of the explosion \citep{jerkstrand2014} \footnote{The nebular spectrum of SN 2011dh has also been modeled by \cite{2013MNRAS.436.3614S} who find $M_{\mathrm{ZAMS}}=$~13--15~\msun\ with an oxygen mass of approximately 0.3~\msun\ at the time of explosion.}.

Lacking spectra of iPTF13bvn in the nebular phase,
we can instead use the $r^{\prime}$-band photometry from the NOT 
obtained at +227~d, +240~d, and +262~d.
Under the (conservative) assumption that all of the $r^{\prime}$-band flux is
due to the [\ion{O}{i}] lines
and with 
$M_{^{56}\rm{Ni}} = 0.05$~\msun, we find 
$O_\mathrm{r}\approx0.018$ at +227~d,
$O_\mathrm{r}\approx0.017$ at +240~d and $O_\mathrm{r}\approx0.015$ at +262~d.  
For SN~2011dh the corresponding values 
are in the range $O_\mathrm{r} \approx 0.012$--0.08, 
as derived from spectral observations between
+200~d to +300~d. For stars with $M_{\mathrm{ZAMS}}$ significantly higher than that of SN~2011dh, 
much higher values for $O_\mathrm{r}$ are predicted; for example, $O_\mathrm{r}\approx0.035$ for $M_{\mathrm{ZAMS}}=17$~\msun\  \citep{jerkstrand2014}. 
Thus, we find
that the late-time oxygen line luminosity of iPTF13bvn is not consistent with
that expected from the  $M_{\mathrm{ZAMS}}=32$~\msun\ progenitor suggested
by \cite{Groh:2013aa}.

\begin{figure}
\centering
\includegraphics[width=\columnwidth]{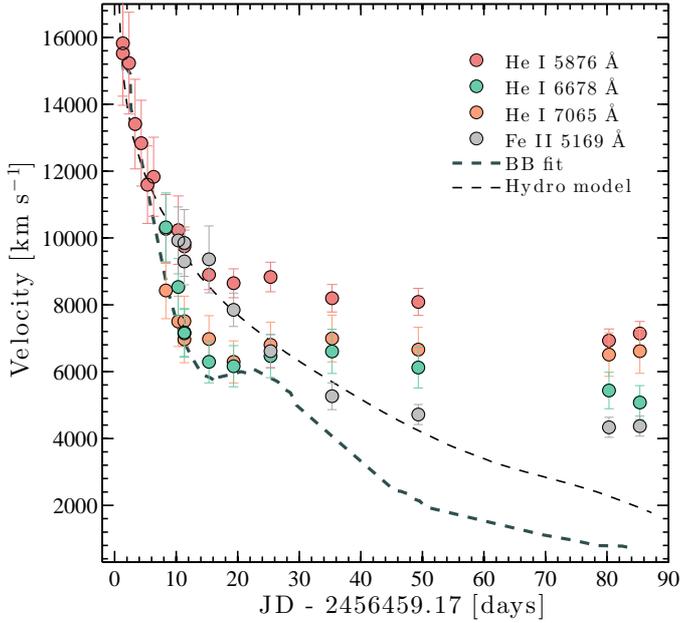}
\caption{Line velocities of iPTF13bvn derived from absorption minima in the spectra (colored circles), the photospheric velocity derived from BB fits to the photometry (thick dashed line) and the photospheric velocity of iPTF13bvn modeled by our hydrodynamical code (thin dashed line).}
\label{fig:line_vel}
\end{figure}

\section{Light curve modeling and progenitor constraints}
\label{sec:hydro}

The semi-analytic framework of \cite{Piro:2012aa}, in combination
with early-time observations,
was used by \cite{2013ApJ...775L...7C} to
constrain the radius of the progenitor of iPTF13bvn to a few solar radii. 
Furthermore, the {\it HST} progenitor candidate
detection and the deduced mass-loss rate by \cite{2013ApJ...775L...7C} led to
the conclusion by both \cite{2013ApJ...775L...7C} and \cite{Groh:2013aa} that the
progenitor of iPTF13bvn was consistent with a compact WR
star. However, in the previous sections we have used semi-analytic
arguments regarding the bolometric LC, 
as well as nebular oxygen emission considerations, 
to show that 
our post-peak data 
are inconsistent with a massive ($>30$~\msun) single-star progenitor. In this section, we attempt to more robustly
constrain some of the properties of the progenitor of iPTF13bvn using a hydrodynamical model for the bolometric LC.

We use our observed multi-band {\it UBVRI} and $g^{\prime} r^{\prime} i^{\prime} z^{\prime}$ LCs of iPTF13bvn, 
the explosion date estimate by
\cite{2013ApJ...775L...7C} as an initial guess for the explosion time, and 
our measured photospheric velocities 
together with a grid of SN models constructed with the hydrodynamical
code \emph{HYDE}, previously used to model the first 100 days of the bolometric LCs of SNe
1993J, 2008ax, and 2011dh (Ergon et al. 2014b, in prep.). The code is
one-dimensional, based on flux-limited diffusion, and follows the
framework described by \cite{1977ApJS...33..515F}\footnote{After the submission of this paper, another paper on iPTF13bvn \citep{2014arXiv1403.7288B} has also appeared. Similar hydrodynamical modeling is performed, showing consistent results.}. The model grid is based on bare He cores evolved until the verge of core collapse using \emph{MESA}\footnote{\href{http://mesa.sourceforge.net}{http://mesa.sourceforge.net}} \citep{2010ascl.soft10083P}. 
For the infrared and ultraviolet regions we use bolometric corrections derived from the bolometric LC of SN~2011dh. When we perform the fits to our model grid, we place equal weights on the diffusion phase of the LC (+1~d to +40~d), the early tail of the LC (+40~d to +100~d), and the
photospheric velocity measurements (+2~d to +40~d).

The best-fitting model, shown in Fig.~\ref{fig:bol_lc} along with the estimated bolometric LC,
constrains the total energy of the explosion to be
$E=0.85^{+0.50}_{-0.40}$$\times$10$^{51}$ erg, 
the synthesized radioactive nickel mass to be
$M_{\Ni}=0.049^{+0.021}_{-0.012}$~\msun, and the total ejecta
mass to be $M_{\mathrm{ej}}=1.94^{+0.50}_{-0.58}$~\msun\ under the assumption
that a 1.5~\msun\ remnant remains at the center of the SN
explosion. The total mass of the He core before the explosion in the model
is thus $M_{\mathrm{He}}=3.44^{+0.50}_{-0.58}$~\msun. The best-fitting explosion time ($t_b$) is 2013 June 15.55 (UT).
The ejecta mass we find is much lower than what is suggested by the
\cite{Groh:2013aa} massive single star progenitor scenario, which
requires\footnote{The total He-core mass of the progenitor proposed by \cite{Groh:2013aa} immediately prior to the explosion is $10.9$~\msun.} $M_{\mathrm{ej}}\approx8$--9~\msun\ assuming a 2--3~\msun\ remnant. We also find that to fit the early rise of the LC up to 10~d past the explosion, our model requires the radioactive 
nickel synthesized in the explosion to be highly mixed out in the ejecta. If the nickel is concentrated toward the center, the diffusion time becomes too large, and the rise of the model LC happens too late to fit the fast rise of the observed early LC.
The reported errors are propagated from the observed quantities; we assume 
an uncertainty in the
distance modulus to NGC 5806 of~$\pm0.3$~mag, an uncertainty in the total
extinction for iPTF13bvn of $\pm0.045$~mag, and uncertainties
in the photospheric velocities of $\pm15\%$. 

We note that our model is generally able to achieve an excellent fit to the observed data, including the peak and the tail of the LC as well as the photospheric velocities\footnote{We do not fit for the photospheric velocities at times later than +40~d. Our model of the photosphere is not applicable at these times.}. Furthermore, if we let the final explosion time vary between the time of the first non-detection until the first detection in the $r^{\prime}$-band, we find that, 
while the quality of the fit is decreased (especially for dates earlier than $t_b-0.25$~d), the high nickel-mixing requirement always remains and the derived explosion parameters are not significantly changed. 
In this sense, our $M_{\Ni}$ and $M_{\mathrm{ej}}$ estimates are quite robust, and we note that they are also very similar to the semi-analytic nickel and ejecta masses derived in Sect.~\ref{sec:Photometry}. 

We also note that the radius constraint of a few solar radii on the progenitor of iPTF13bvn by \cite{2013ApJ...775L...7C} was partly based on assumed typical explosion parameters for a SN~Ib. Our model does not directly constrain the radius; it is constant across the models in our grid. However, the explosion parameters we derive permit a larger radius of the progenitor within the \cite{Piro:2012aa} semi-analytic framework.

\section{Summary and conclusions}

We have confirmed the classification of iPTF13bvn as a Type Ib SN, and also
that the pre-explosion {\it HST} progenitor identification suggested
by \cite{2013ApJ...775L...7C} is very likely correct
(Sect.~\ref{sec:prog_13bvn}). 
While the color and
luminosity of the progenitor in the {\it HST} images are consistent with
those of a WR star, we have shown that the later-time photometry
(Sect.~\ref{sec:Photometry}) 
of iPTF13bvn is not consistent with a
single, very massive WR progenitor scenario. Our hydrodynamical model
can fit the observed bolometric light curve to constrain the 
total ejecta mass to be  $\sim1.9$~\msun. 
The corresponding mass of
the He core of the progenitor is $\sim 3.4$~\msun\ immediately prior to the
explosion. The synthesized nickel mass is constrained 
to be $\sim0.05$~\msun. 
The model also requires the synthesized nickel to be highly mixed out in the ejecta, consistent with the
high average mixing found for the sample of SNe~Ibc by \cite{taddia2014}. 
The total ejecta mass and our limits on the late-time
oxygen emission are both inconsistent with what would be expected from
a single massive WR progenitor with $M_{\mathrm{ZAMS}} \approx 30$~\msun\ as
suggested by \cite{Groh:2013aa}. We note that while it could be argued that a significant part of the He core could fall back onto the central compact object, such a scenario that still produces an apparently nickel-powered LC, normal observed photospheric velocities, and the measured oxygen emission appears unlikely. 
We also note that while the deduced high mass-loss rate was interpreted by 
\cite{Groh:2013aa} as favoring a massive WR star, the mass-loss estimate depends on an assumed high stellar wind velocity. A lower wind velocity in combination with the radio measurements of \cite{2013ApJ...775L...7C} can also be consistent with a less massive star.

To date, the SN~Ibc showing the largest deviation in LC shape
compared to the typical SN~Ibc LC \citep[e.g.,][]{2011ApJ...741...97D,taddia2014}  
is the bolometric LC of SN~Ic~2011bm  presented by
\cite{Valenti:2012aa}. SN~2011bm has a very high peak luminosity and
(most importantly) a very slow evolution of the bolometric LC and a
broad peak, which is difficult to explain without
assuming a very high ($> 30$~\msun) main-sequence mass for the
progenitor. This is the kind of SN LC we would expect from 
a scenario including the very massive progenitor star suggested by 
\cite{Groh:2013aa}.
It is clear that the bolometric LC of iPTF13bvn is very
different from that of SN~2011bm. 

If the progenitor is less massive, a binary system
for the progenitor is perhaps the most natural conclusion. Using the
progenitor constraints from our hydrodynamical model, the evolutionary
models for binary systems by \cite{Yoon:2010aa} predict values for the
radius, mass-loss rate, and hydrogen content in the progenitor that
are all consistent with what can be derived from the early-time
observations by \cite{2013ApJ...775L...7C}. In contrast, current single-star
evolutionary synthesis models \citep[e.g.,][]{Groh:2013ab} have a harder
time producing a progenitor with a sufficiently low mass to match the
observed properties; stars with low enough ZAMS masses to explain the low ejecta mass, as well as the observed nebular oxygen emission constraints, are not
predicted to produce SNe~Ibc in this context. 

The ultimate test to assess the nature of the progenitor will be to
reobserve the location of iPTF13bvn with {\it HST} after the SN has
faded. A single massive WR progenitor scenario predicts that
the progenitor candidate would completely disappear, while a less
massive progenitor in a binary system would lead to a smaller 
decrease in the luminosity at
the location of the progenitor. 

Finally, we emphasize that the use of the new \cite{jerkstrand2014} nebular modeling in conjunction with photometry
during the nebular phase could be an efficient way to constrain
the ZAMS masses for a larger sample of stripped-envelope CC~SNe. Future studies of nebular spectra of iPTF13bvn are planned, and we are also performing a detailed investigation of the host environment of the SN (Fremling et al. 2014b, in prep.). 

\begin{acknowledgements}

The Oskar Klein Centre is funded by the Swedish Research Council.
This work is partially based on observations made with the Nordic Optical Telescope, operated by the Nordic Optical Telescope Scientific Association at the Observatorio del Roque de los Muchachos, La Palma, Spain, of the Instituto de Astrofisica de Canarias. The data presented here were obtained in part with ALFOSC, which is provided by the Instituto de Astrofisica de Andalucia (IAA) under a joint agreement with the University of Copenhagen and NOTSA.
Some of the data presented herein were obtained at the W. M. Keck
Observatory, which is operated as a scientific partnership among the
California Institute of Technology, the University of California, and
NASA; the observatory was made possible by the generous financial
support of the W. M. Keck Foundation. A. G.-Y. is supported by the EU/FP7 via ERC grant 307260, ``The Quantum Universe'' I-Core program by the Israeli Committee for planning and budgeting, by ISF,
GIF, and Minerva grants, and by the Kimmel award.
A. V. F.'s group at UC Berkeley has received generous financial assistance
from the Christopher R. Redlich Fund, the TABASGO Foundation,
NSF grant AST-1211916, and NASA grants
AR-12623 and AR-12850 from the
Space Telescope Science Institute (which is operated by AURA,
Inc., under NASA contract NAS 5-26555).
We extend our thanks to the following people for their various contributions to this work: Shri Kulkarni, Mansi M. Kasliwal, Ofer Yaron, Paul Vreeswijk, Daniel Perley, Joel Johansson, Anders Jerkstrand, Kelsey Clubb, Ori Fox, Patrick Kelly, Barak Zackay, Adam Waszczak, Donald O'Sullivan, and Thomas Augusteijn.

\end{acknowledgements}

 \bibliography{13bvn_12os}

\end{document}